# Hollow carbon sphere/metal oxide nanocomposite anodes for lithium-ion batteries

K. Wenelska[1,*], A. Ottmann[2,*], P. Schneider[2], E. Thauer[2], R. Klingeler[2,3], E. Mijowska[1]

[1]Institute of Chemical and Environment Engineering, West Pomeranian University of Technology, ul. Pulaskiego 10, 70-322, Szczecin, Poland
[2]Kirchhoff-Institute for Physics, Heidelberg University, INF 227, 69120 Heidelberg, Germany
[3]Centre for Advanced Materials, Heidelberg University, INF 225, 69120 Heidelberg, Germany

*Both authors contributed equally to this work.

## Abstract

Hollow carbon spheres (HCS) covered with metal oxide nanoparticles ($SnO_2$ and $MnO_2$, respectively) were successfully synthesized and investigated regarding their potential as anode materials for lithium-ion batteries. Raman spectroscopy shows a high degree of graphitization for the HCS host structure. The mesoporous nature of the nanocomposites is confirmed by Brunauer-Emmett-Teller analysis. For both metal oxides under study, the metal oxide functionalization of HCS yields a significant increase of electrochemical performance. The charge capacity of $HCS/SnO_2$ is 370 mAh/g after 45 cycles (266 mAh/g in $HCS/MnO_2$) which clearly exceeds the value of 188 mAh/g in pristine HCS. Remarkably, the data imply excellent long term cycling stability after 100 cycles in both cases. The results hence show that mesoporous HCS/metal oxide nanocomposites enable exploiting the potential of metal oxide anode materials in Lithium-ion batteries by providing a HCS host structure which is both conductive and stable enough to accommodate big volume change effects.

**Introduction**

Rechargeable Li-ion batteries (LIBs) are currently the dominant energy storage technology for portable electronic devices. Despite their enormous commercial success in this field, for more demanding high power and high energy applications new materials have to be developed which enable significantly improved specific energy densities, capacities, and power rates in combination with long life time and cycling stability [1]. In case of the anodes, the most commonly used reference is graphite, which is a low-cost material with good cyclic stability and low electrochemical potential. Its limited reversible intercalation capacity (theoretically 372 mAh/g) and poor rate performance hinder its application in high-performance LIBs [2]. Accordingly, extensive efforts have been done to develop new high-performance anode materials for next-generation LIBs. One promising approach aims at utilizing redox-active metal oxides such as $Fe_2O_3$ [3], $Fe_3O_4$ [4], $SnO_2$ [5], and $Co_3O_4$ [6] which exhibit very high theoretical specific capacities as they are capable of converting up to 6 Li per formula unit. Despite significant progress, however, metal oxide-based anodes do not yet reach their full potential which is mainly caused by their low conductivity and by large volume changes during dis-/charge cycling [7]. In particular, the large strain upon electrochemical cycling causes disintegration of the active material into small particles which is accompanied by significant losses of reversible capacity [8].

A materials-science based answer to these issues applies nanoscaled metal oxides embedded into conductive carbon structures. Such nanocomposites have indeed emerged as a promising method towards high-performance anode materials [9,10]. On the one hand, downscaling the active material yields short transport lengths for both, electrons and Li-ions, higher specific surface areas, and better accommodation of strain upon Li insertion/extraction [11]. To date, such composites are commonly prepared by simply coating pyrolytic carbon species on nanoscaled oxides [12,13]. However, while increasing the overall electronic conductivity of the materials, pyrolytic carbon does not provide an effective host structure for accommodating the strain caused by large volume expansion and such materials hence offer only limited cyclic stability. Alternatively, creating composites with porous characteristics has been proposed to overcome these limitations [14–17]. In this regard, hollow carbon spheres (HCS) can provide a host structure which is both conductive and stable enough to accommodate big volume change effects. Those benefits have already been utilized in composites with metal oxides such as $SnO_2$ and $Co_3O_4$, which show very promising electrochemical properties [18–20]. Here, we report a facile impregnation method to synthesize hollow carbon spheres (HCS) [21] covered with $SnO_2$ and $MnO_2$ nanoparticles as advanced anode material for high performance LIBs (see Fig. 1). The resulting nanocomposite materials display reversible capacities larger than pristine HCS as well as good cycling performance.

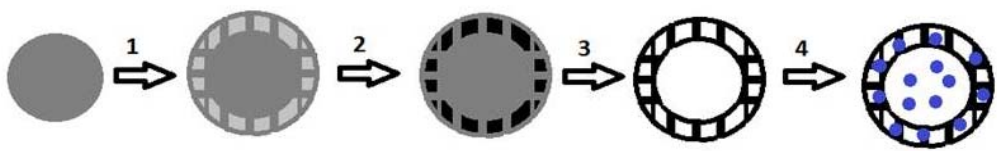

Fig. 1: Synthesis schematic of hollow carbon spheres (black) decorated with metal oxide nanoparticles (blue). Mesoporous silica spheres (grey) are being carbonized (step 2). After removal of silica (step 3), the resulting hollow carbon sphere is functionalized by metal oxide nanoparticles (step 4). (See the text.)

**Experimental section**.

**Synthesis of solid $SiO_2$ nanospheres ($SiO_2$).** $SiO_2$ spheres were prepared in a modified Stoeber process [22]. Tetraethyl orthosilicate (TEOS) (1.5 ml) was added to a mixture of ethanol (50 ml) and concentrated ammonia (28 wt%, 2.5 ml). Then the solution was stirred for 24 h. Afterwards, the product was separated by filtration, washed with ethanol and dried [21].

**Synthesis of mesoporous silica spheres ($SiO_2$@m-$SiO_2$).** In a typical synthesis, 100 mg of $SiO_2$ was dispersed in a solution containing cetyltrimethylammonium bromide (CTAB, 0.8 g), deionized water, concentrated ammonia (28 wt%), and 60 ml ethanol. The suspension was sonicated and stirred for 60 min, then 1.43 ml of TEOS was added dropwise while stirring. The suspension was stirred for another 6 h, and afterwards, the product was filtered, washed with ethanol and water several times. Finally, the sample was dried in air at 100 °C for 24 h.

**Carbonization of $SiO_2$@m-SiO2 and removal of silica (HCS)**. The dried SiO2@m-SiO2_CTAB spheres were used as a template to prepare the hollow mesoporous carbon spheres using chemical vapor deposition (CVD). The silica spheres were placed in an alumina boat and put into a tube furnace. Argon and ethylene were introduced at a flow rate of 100 and 30 sccm, respectively. The temperature was raised to 800 °C, and the CVD reaction time was 3 h. Afterwards, the resulting SiO2@m-SiO2_C spheres were thoroughly washed with hydrofluoric acid to remove the silica components and finally hollow carbon spheres (HCS) were obtained.

**Functionalization of HCS with metal oxide nanoparticles.** Two samples of HCS modified by metal oxide nanoparticles (HCS/$SnO_2$ and HCS/$MnO_2$, respectively) were prepared according to the following procedure: 150 mg of HCS and 150 mg manganese acetonate (product referred to as HCS/$MnO_2$) or tin chloride (product referred to as HCS/$SnO_2$) were dispersed in 250 ml of ethanol and sonicated for 2 h. Afterwards, the mixture was stirred for another 24 h. Finally, the product was dried in air at 100 °C for 24 h.

**Characterization.** The morphology of the samples was investigated by a FEI Tecnai F30 transmission electron microscope (TEM) with a field emission gun operating at 200 kV and Energy-dispersive X-ray spectroscopy (EDX) as one mode. Powder X-ray diffraction (XRD) was performed on a Philips diffractometer using Cu-$K_\alpha$ radiation. Raman scattering was studied on a Renishaw micro-Raman spectrometer (λ = 720 nm). $N_2$ adsorption/desorption isotherms were accquired at liquid nitrogen temperature (77 K) using a Micromeritics ASAP 2010 M instrument, and the specific surface area was calculated by the Brunauer Emmett Teller (BET) method. Thermogravimetric analysis (TGA) was carried out on 10 mg samples using the TA SDT-Q600 at a heating rate of 10 °C/min from room temperature to 900 °C in air flow (100 ml/min).

Electrochemical studies by means of cyclic voltammetry and galvanostatic cycling were carried out in Swagelok-type two-electrode cells by a *VMP3* (Bio-Logic) potentiostat (see Ref. [23]). The working electrodes were prepared from a mixture of pristine material with carbon black (*Super P*, Timcal) and polyvinylidene fluoride (PVDF) binder (Solvay Plastics) in a weight ratio of 70:15:15. Additional carbon black was added in order to assure the mechanical stability of the electrodes. PVDF was dissolved in N-Methyl-2-pyrrolidon (NMP) and subsequently the active material and carbon black were mixed with the solution. The resulting slurry was pasted on circular copper plates and dried at 100 °C in a vacuum furnace (< 5 mbar) over night. After mechanical pressing at 10 MPa, the electrodes were dried again. The Swagelok-type cells were assembled in an argon atmosphere glovebox ($H_2O$, $O_2$ < 0.5 ppm) with the working electrode, a lithium metal (Alfa Aesar) counter electrode, which had been pressed on a circular nickel plate, and two layers of glass microfibre separator (*GF/D*, Whatman). 200 μl of a 1M solution of $LiPF_6$ in 1:1 ethylene carbonate (EC) and dimethyl carbonate (DMC) was used as electrolyte (*LP30,* Merck). While measuring, the cells were held at 25 °C in a climate chamber. For the CVs, the scan rate was fixed at 0.1 mV/s in the voltage range of 0.01 - 3.00 V and GPCL measurements were done at current densities of 100 – 1000 mA/g.

**Results and discussion**

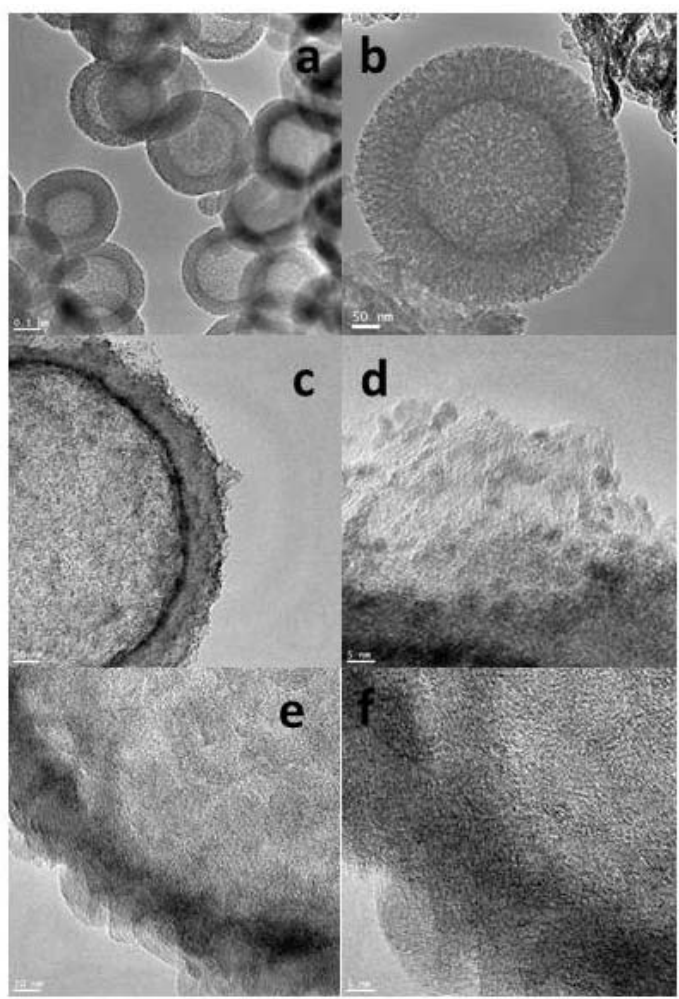

Fig. 2: TEM images of hollow carbon spheres (a,b), HCS/SnO$_2$ (c,d), HCS/MnO$_2$ (e,f).

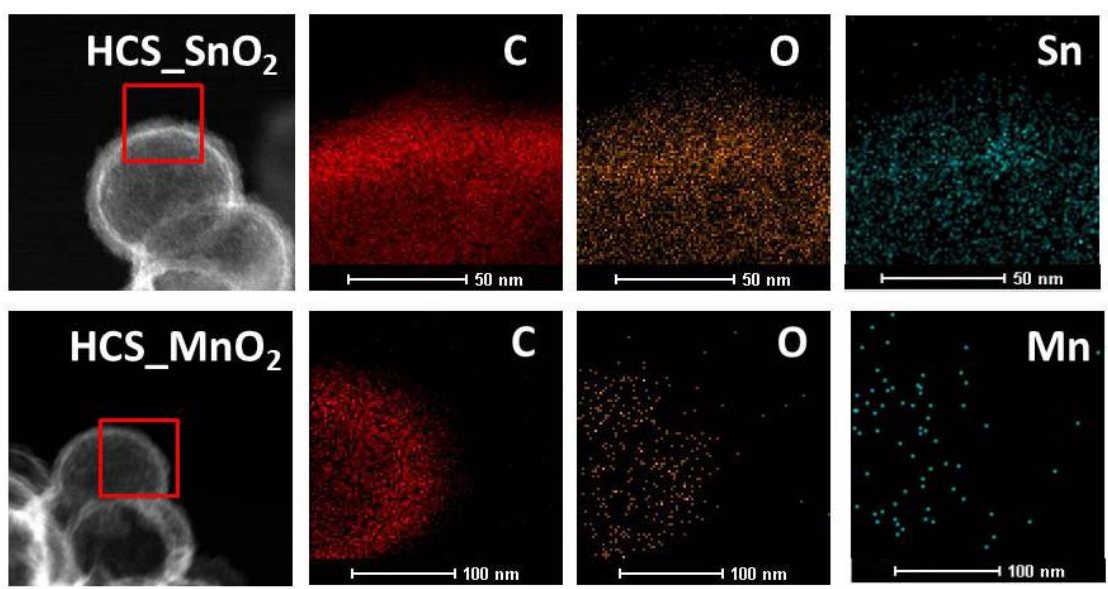

Fig. 3: EDS mapping of HCS/SnO$_2$ and HCS/MnO$_2$ (see the text).

Fig. 2 shows TEM images of pristine hollow carbon spheres and of HCS/metal oxide composites with different metal oxide loading. As observed in Fig. 2(a,b), the pristine HCS are uniform in diameter which amounts to about 250 nm. The shell thickness is about 90 nm. Functionalization with $SnO_2$ nanoparticles yields HCS/$SnO_2$ with a small metal oxide particle size distribution ranging from 3 to 5 nm. The particle sizes have been derived from averaging over 100 nanoparticles observed in TEM images (Fig. 2(c,d)). The $SnO_2$ nanoparticles are distributed on the surface of HCS homogeneously. Similar results are obtained for HCS/$MnO_2$ (Fig. 2(e, f)). However, the diameter of $MnO_2$ nanoparticles is smaller, ranging from 1 to 3 nm. As shown in Fig. 3, EDX elemental mapping clearly reveals that the elements Sn, O, and C are evenly distributed throughout the HCS/$SnO_2$ nanocomposite. Similarly, the elemental mappings of HCS/$MnO_2$ indicate the presence of Mn, O, and C. All detected elements seem to be rather homogeneously distributed in the sample. In summary, the EDS data clearly show that $SnO_2$ and $MnO_2$ are located both in the core and the shell of the hollow carbon spheres.

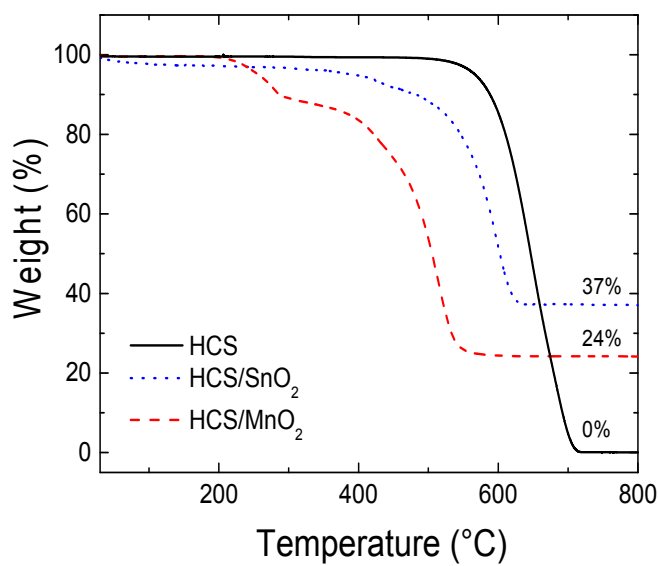

Fig. 4: TGA profiles of HCS, HCS/$SnO_2$, and HCS/$MnO_2$.

The carbon content and the quality of the materials were evaluated by means of thermogravimetric analysis (TGA) (Fig. 4) and Raman spectroscopy (Fig. 5). The thermogravimetric analyses of the HCS, HCS/$SnO_2$, and HCS/$MnO_2$ samples present the thermal stability by monitoring the change of weight during heating. According to the TGA results (Fig. 4), the pristine HCS start to oxidize around 550 °C. The HCS are completely exhausted when the temperature is increased to ~700 °C in air, and the ash content of 0 wt% indicates high purity of the HCS [21]. TGA measurements for HCS with metal oxide nanoparticles show ash contents of 24 wt% and 37 wt% for $MnO_2$ and $SnO_2$, respectively. In comparison to the pristine HCS, the stabilities of both metal oxide-functionalized HCS materials are weaker. The main weight loss starts at around 400 °C for both HCS/metal oxide materials. One may conclude that the interaction of the metal oxides and the carbon induces lower stability of HCS. Additionally, a mass loss of ~10% is observed in the thermal profile of HCS/MnO$_2$ starting at ~210°C. This feature presumably originates from the decomposition of residual manganese acetate.

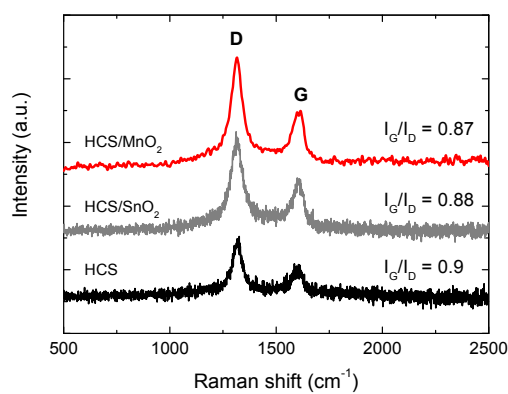

Fig. 5: Raman spectra of HCS, HCS/SnO$_2$, and HCS/MnO$_2$.

In the Raman spectra, two distinct peaks are detected around 1314 and 1595 cm$^{-1}$, which correspond to disordered carbon (D) and ordered graphitic carbon (G), respectively, as shown in Fig. 5. In case of the pristine HCS, the $I_G/I_D$ intensity ratio of the G- and D-lines amounts to 0.9, indicating a high degree of graphitization. Note, that such a high crystallinity in general supports high electrical conductivity as desirable for application in Li-ion electrodes. Upon deposition of the metal oxide nanoparticles, the relation of G- to D-band intensities slightly decreases which is consistent with the assumption that additional defects are formed in the HCS structures. The D-mode is most pronounced for HCS/MnO$_2$, indicating the highest concentration of defects and hence the lowest stability as it is experimentally observed in the TGA data.

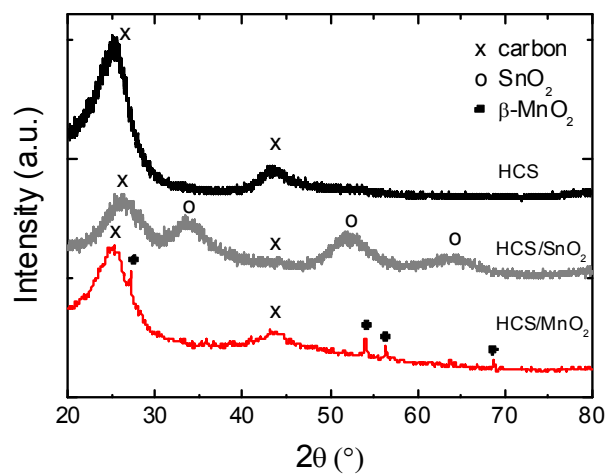

Fig. 6: XRD spectra of HCS, HCS/SnO$_2$, and HCS/MnO$_2$. The crosses, open and filled cycles label diffraction peaks associated with graphitic carbon, SnO$_2$, and β-MnO$_2$, respectively.

The XRD patterns of the synthesized samples confirm the presence of graphitic carbon and metal oxides in the functionalized materials (Fig. 6). For pristine HCS, there are two broad peaks at 24.9° and 42° which can be ascribed to graphitic carbon. In addition to the carbon peaks, HCS/metal oxides exhibit further diffraction peaks. In HCS/SnO$_2$, there are major peaks at 33.9°, 51.8°, and 65.8°, which can be attributed to tetragonal SnO$_2$ nanoparticles. The XRD pattern of HCS/MnO$_2$ shows characteristic peaks for β-MnO$_2$ at 2Θ = 27°, 54°, 56°, 68.7°.

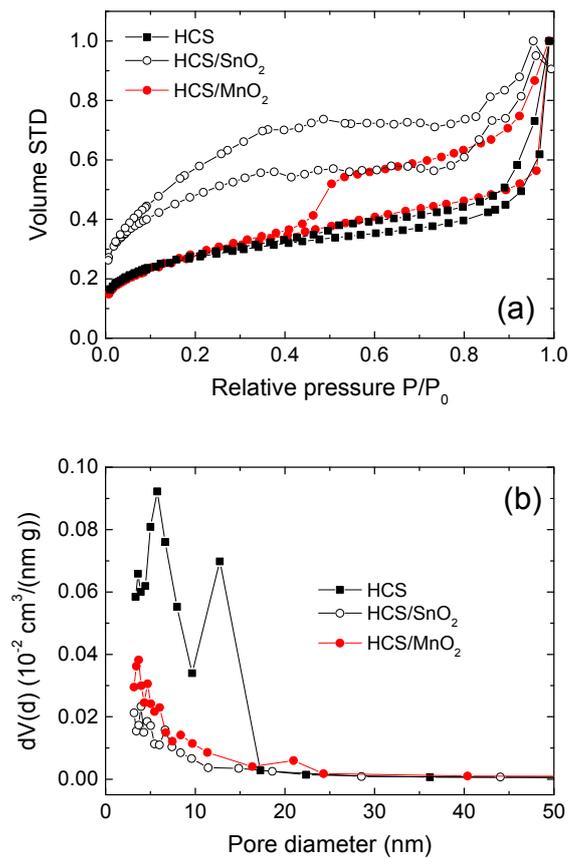

Fig. 7: (a) $N_2$ adsorption/desorption isotherms and (b) pore size distribution profile for HCS, HCS/SnO$_2$, and HCS/MnO$_2$.

The Brunauer−Emmett−Teller (BET) specific surface area and the porosity of all samples studied by nitrogen adsorption−desorption analysis confirms the mesoporous nature of the materials (Fig. 7(a)). The data show typical type IV behavior with a H4 hysteresis loop, indicating a disordered mesoporous structure [24]. Quantitatively, the BET specific surface area of pristine HCS amounts to 652 m$^2$/g. It significantly drops for the metal oxide covered nanospheres to 302 m$^2$/g for HCS/SnO$_2$ and 184 m$^2$/g for HCS/MnO$_2$, respectively. The corresponding pore size distribution in HCS calculated by means of the BET method from the adsorption branch reveals non-uniform pores centered at approximately 3.6, 6.0, and 12.7 nm (Fig. 7(b)). In HCS/SnO$_2$, the pores sizes are 3.3, 4.1 and 6.7 nm, and in HCS/MnO$_2$ they are centered at 3.7, 4.7, 6.1, 8.4 and 21 nm. The observation of smallest pore volume and surface area in HCS/MnO$_2$ suggests that the MnO$_2$ nanoparticles block the pores of the hollow carbon spheres in this case.

**Electrochemical properties**

The cyclic voltammograms (CV) in Fig. 8 show the characteristics of the electrochemical processes upon variation of the potential with 0.1 mV/s in the range between 0.01 and 3.0 V. For HCS, several well-known features in carbon structures are observed. The first cycle, starting with a negative voltage ramp at 3.1 V, shows reduction peaks at 0.01 V, 0.6 V and 1.35 V. Oxidation occurs at 0.2 V, around 1.2 V, and above 2 V. The reduction peak at 0.6 V can be attributed to the formation of a passivating solid electrolyte interface (SEI) on the carbon surfaces [25]. Correspondingly, this peak disappears upon further cycling but in the overall behavior a shoulder shows up around 0.8 V (see Fig. 8(b)). It might indicate an ongoing irreversible contribution from the SEI formation or might be related to the oxidation process at 1.2 V which decreases upon cycling as well. In contrast, the redox pair at 0.01/0.2 V corresponds to reversible de-/intercalation of Li-ions into the carbon structures. With increasing cycle number the peak height of the reduction peak decreases while the oxidation peak intensity increases and shifts to slightly lower potentials. The origin of the reduction peak appearing at 1.35 V in the first cycle only is unknown. The oxidative contributions above 2 V are presumingly originating from the cell setup.

The evolution of the dis-/charge capacities in HCS (Fig. 9) upon cycling is typical for carbon structures. In the initial charge/discharge process performed at 100 mA/g the material shows capacities of 1305 mAh/g and 269 mAh/g, respectively. The next cycles show a huge drop-off mainly due to the absence of the irreversible contribution of SEI formation in the first cycle. Quantitatively, the discharge capacity amounts to 190 mAh/g after 10 cycles. This is clearly below the theoretical capacity of graphite (372 mAh/g). We associate this observation to the presence of amorphous carbon as indicated by the Raman and the XRD data (Figs. 5 and 6) which implies lower capacities as compared to graphite [26]. Increasing the charge/discharge current to 250, 500 and 1000 mA/g yields 153/149 mAh/g (after 20 cycles), 124/123 mAh/g (30 cycles), and 103/103 mAh/g (40 cyles). In cycle 45, back at 100 mA/g, the charge and discharge capacities are 188 mAh/g and 179 mAh/g, respectively, which is more than 90% of the capacities reached in the 10$^{th}$ cycle.

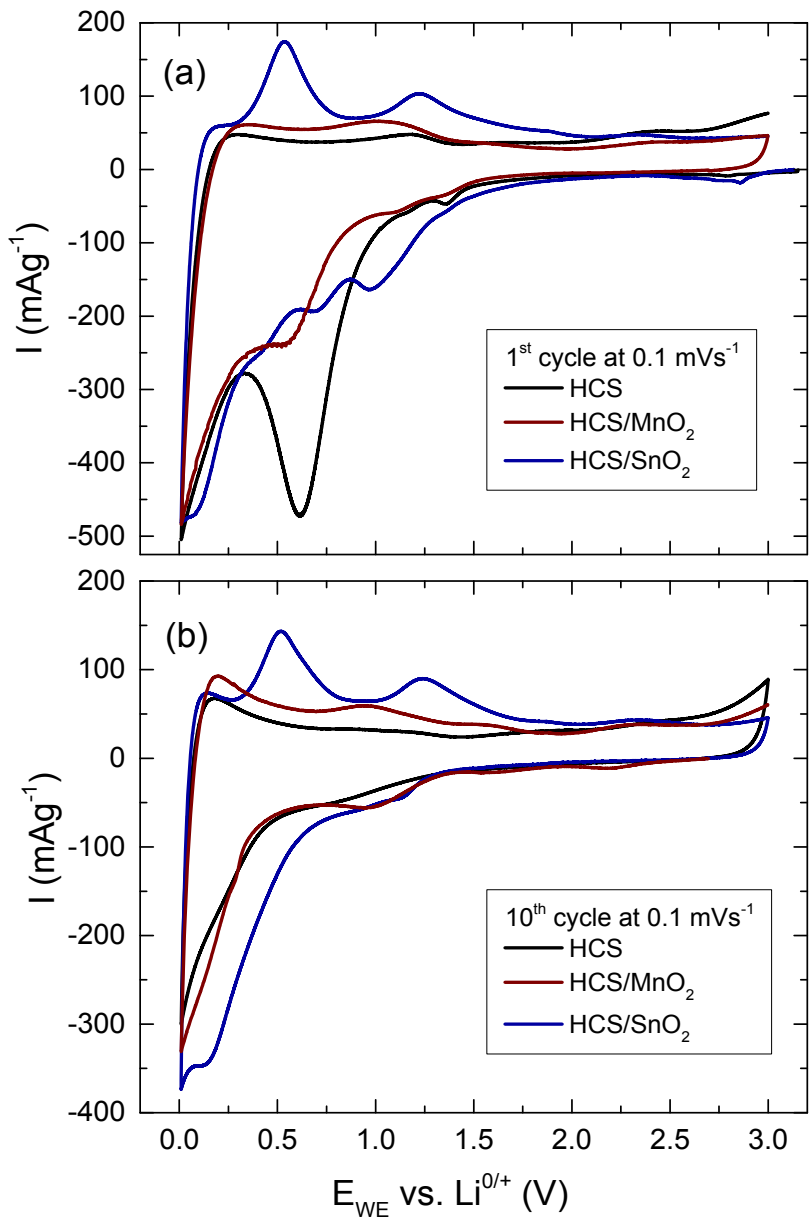

Fig. 8: Cyclic voltamograms of HCS, HCS/SnO$_2$, and HCS/MnO$_2$, obtained at a scan rate of 0.1 mV/s. (a) shows the 1$^{st}$ cycle, (b) the 10$^{th}$ cycle.

Functionalization of HCS by means of SnO$_2$ and MnO$_2$ nanoparticles yields additional features in the CVs. In cycle 1 (Fig. 8(a)), initial reduction processes of SnO$_2$ and MnO$_2$ to corresponding metals and formation of amorphous Li$_2$O shows up. In case of HCS/SnO$_2$, in addition to the processes for HCS discussed above, there are reduction peaks at 0.97 V, 0.69 V, 0.42 V, and 0.07 V, while additional/more pronounced oxidations are observed around 0.53 V and 1.23 V (the latter being much more pronounced than for HCS). Upon further cycling, the three reduction peaks at 0.97 V, 0.69 V, and 0.42 V merge to a plateau-like shoulder.

According to [27,28] the electrochemical behavior of SnO2 can be described via

$$4\ Li^+ + SnO_2 + 4e^- \leftrightarrows Sn + 2\ Li_2O \qquad (1)$$

$$x\ Li^+ + Sn + xe^- \leftrightarrows Li_xSn\ (0 \leq x \leq 4.4) \qquad (2)$$

The redox pair at 0.07/0.53 V can be attributed to the reversible alloying and dealloying of Sn and Li (reaction (2)) [19]. The reduction peak at 0.42 V might also be part of this alloying process [29]. The reduction peaks at 0.69 V and 0.97 V in the first cycle presumingly correspond to the reduction of $SnO_2$ to $Sn+Li_2O$ (reaction 1) and to the SEI formation [19,29,30]. The corresponding oxidation of Sn is supposed to appear around 1.2 V, i.e. this process superimposes another oxidation peak around 1.2 V which is also present in the case of pristine HCS. It is worth mentioning that the specific peak current in the voltage range of SEI formation is just ~200 mA/g compared to ~500 mA/g for HCS. We attribute this diminishment to $SnO_2$ nanoparticles on the HCS surface which alter the surface area and the chemistry of the interface to the electrolyte.

The presence of $SnO_2$ nanoparticles enables alloying of Sn (reaction (2)), thereby increasing the capacity of pure HCS. Indeed, the charge/discharge capacities (370 mAh/g and 364 mAh/g) in the 45$^{th}$ cycle are much higher than the respective values of pure HCS (188 mAh/g and 179 mAh/g). Quantitatively, the experimental values are slightly lower than if the full theoretical capacity of the completely reversible process (2) is considered. A maximum value of $x = 4.4$ corresponds to a theoretical capacity of 783 mAh/g. Considering the mass fraction of $SnO_2$ being 37% (and 63% of HCS with 179 mAh/g in cycle no. 45), a completely reversible process would result in a specific discharge capacity of 402 mAh/g. The observed value of 364 mAh/g hence would indicate $x \sim 3.8$ in cycle no. 45, i.e. rather small irreversibility. We note, however, that the $SnO_2$ coating also may diminish the available capacity of HCS as compared to the pure reference material due to blocking of pores for instance, which is indicated by a lower specific surface area of the HCS/$SnO_2$ composite (Fig. 7). We emphasize that rather large charge/discharge capacities of 315/308 mAh/g are still achieved after 100 cycles, and therefore 85% of the capacity is maintained over 55 cycles at 100 mA/g (from 45$^{th}$ to 100$^{th}$).

Since HCS/$MnO_2$ exhibits a lower fraction of $MnO_2$, the effect of metal oxide functionalization is less pronounced than in HCS/$SnO_2$. The conversion reaction is however visible in the CVs in Fig. 8 as demonstrated by the broad oxidation peak around 1 V and a reductive feature around 0.35 V, which is overlain in the 1$^{st}$ cycle and hardly visible in subsequent ones. This redox pair is associated to the following reaction [8,31]:

$$MnO_2 + 4\ Li^+ + 4e^- \leftrightarrows 2\ Li_2O + Mn \qquad (3)$$

Comparing the peak current of the SEI formation in HCS and HCS/$MnO_2$, i.e. 500 mA/g and 250 mA/g, respectively, illustrates lower activity of the functionalized material. Qualitatively, this corresponds to the same effect as in the case of HCS/SnO2, namely a significantly reduced surface area of nearly 4 times less, i.e. a reduced interfacial area of carbon to the electrolyte.

The charge/discharge capacity in the 45$^{th}$ cycle is 266/257 mAh/g compared to 188/179 mAh/g for HCS. We conclude that the present MnO$_2$ mass fraction of 24% is associated to a discharge capacity of 121 mAh/g which is less than half of the theoretical value of 296 mAh/g (0.24 * 1233 mAh/g). A likely reason for the inferior efficiency compared to the HCS/SnO$_2$ sample is the smaller specific surface area (184 vs. 302 m$^2$/g) and smaller pore volumes which might e.g. hinder efficient Li$^+$ transport. Furthermore, residual precursor (manganese acetate) in the material, as indicated by the TGA result (Fig. 4), would also have a considerably negative impact on the electrochemical performance because it reduces the active material's fraction. Nevertheless, capacity retention of more than 90% over 55 cycles (from 45$^{th}$ to 100$^{th}$) at 100 mA/g is observed. This cyclic stability (similar for HCS/SnO$_2$) argues for the benefit of utilizing HCS in order to maximize the potential of metal oxide anode materials by strain accommodation and by providing a conductive network.

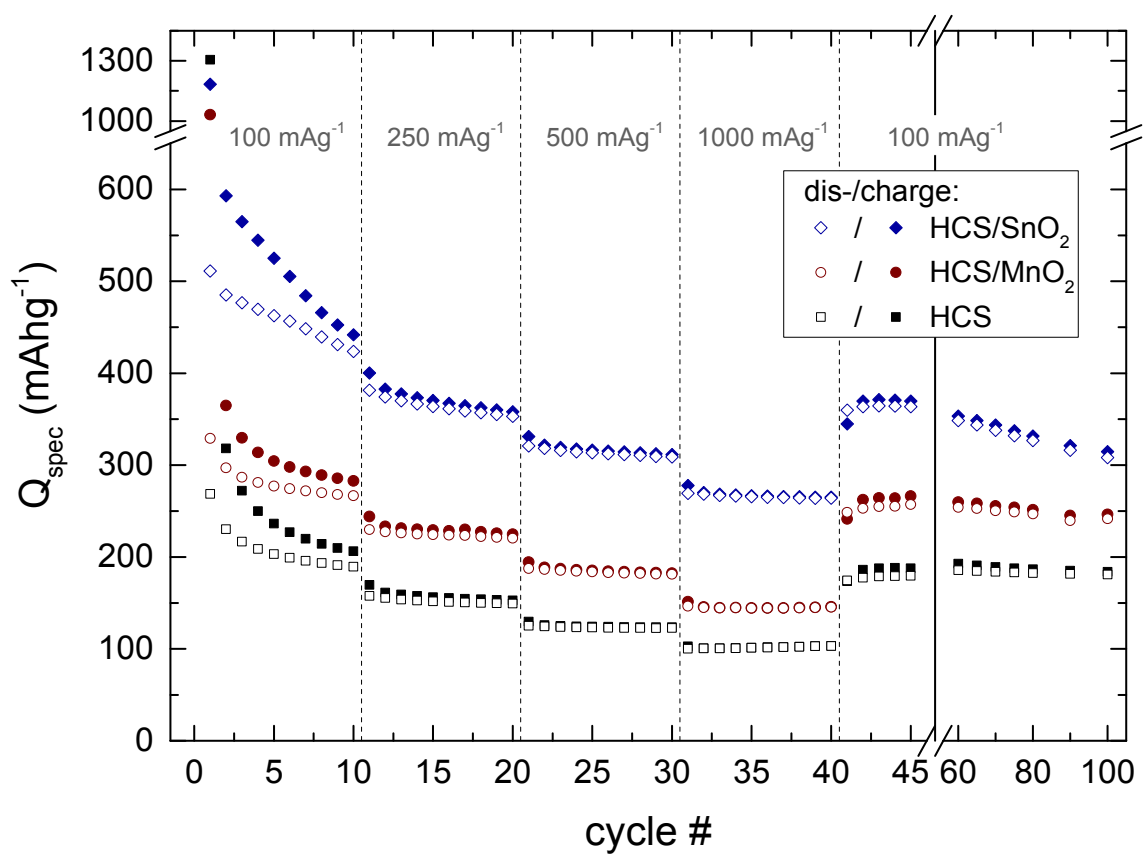

Fig. 9: Gravimetric specific capacities vs. cycle number of HCS, HCS/SnO$_2$, and HCS/MnO$_2$, at different scan rates.

**Summary and conclusions**

In summary, we propose a strategy to produce a new anode material for Li-ion batteries by embedding metal oxide nanoparticles ($SnO_2$, $MnO_2$) in a composite with hollow carbon nanospheres. The intensity ratio of the G- and D-lines observed in Raman spectroscopy indicates a reasonably high degree of graphitization in all materials. Upon functionalization, however, the BET specific surface area of pristine HCS of 652 $m^2$/g decreases for the metal oxide covered nanospheres to 302 $m^2$/g for HCS/$SnO_2$ and 184 $m^2$/g for HCS/$MnO_2$, respectively. While the mesoporous nature of the resulting HCS/metal oxide composites is confirmed by BET analysis our data suggest that in HCS/$MnO_2$ the metal nanoparticles block the pores of the hollow carbon spheres. For both metal oxides under study, functionalization of HCS yields a significant increase of capacity as compared to the pristine material. To be specific, the discharge capacity of HCS/$SnO_2$ is 364 mAh/g after 45 cycles with dis-/charge currents from 100 to 1000 mA/g which clearly exceeds the value of 179 mAh/g in pristine HCS. Remarkably, the long term cycling stability shows a large discharge capacity of 308 mAh/g after 100 cycles. In contrast, HCS/$MnO_2$ shows less enhancement of the electrochemical performance as it reaches 257 mAh/g in cycle no. 45. This smaller enhancement compared to HCS/$SnO_2$ may be attributed to a significantly smaller specific surface area. However, the general strategy of assembling a mesoporous HCS/metal oxide material in order to maximize the potential of metal oxide anode materials is confirmed as excellent long term cycling behavior is observed in HCS/$MnO_2$ as well.

The present work hence reveals advantages of the lithium storage in HCS/metal oxide materials providing a route to develop high-performance mesoporous hybrid materials. Considering that metal oxides are cost-effective and a feasible large-scale production of HCS, the synthesized materials hold great potential for real applications. This effective strategy can be easily expanded to construct other high-performance architectures of functionalized HCS with other metal oxides, providing a general and effective approach towards high-performance metal-oxide- based anodes. In addition, the synthetic route developed in this work is facile and easily extendable to other Li-storable metals or alloys for the development of advanced anode materials of Li-ion batteries.

**Acknowledgments**

The authors are grateful for financial support of German Academic Exchange Service DAAD (PPP project no. 56269175), the NanoTech-Initiative of the Baden-Württemberg-Stiftung (Project CT3: Nanostorage) and the National Science Centre SONATA BIS - UMO-2012/07/E/ST8/01702. A.O. acknowledges support by the IMPRS-QD.